\documentclass[journal,twoside]{IEEEtran}

\pdfoutput=1
\usepackage[nocompress]{cite}
\usepackage{color}
\usepackage{amsfonts,amssymb}
\usepackage{amsmath,bm}
\usepackage{ifpdf}
\usepackage{epsfig}
\usepackage[small]{subfigure}
\usepackage{fmtcount}
\usepackage[T1]{fontenc}
\usepackage{balance}
\usepackage{xcolor}
\usepackage{multirow}

\usepackage{perpage} 
\MakePerPage{footnote} 

\usepackage{array}
\usepackage{eqparbox}
\usepackage{color}
\usepackage{graphicx} 
\usepackage{url} 
\usepackage[small]{subfigure}
\usepackage{amsmath}
\usepackage{makecell}
\usepackage{balance}
\usepackage{tabu}
\usepackage{algorithm}
\usepackage{algorithmic}
\usepackage{longtable}

\usepackage{pifont}
\usepackage[utf8]{inputenc}
\usepackage[T1]{fontenc}

\usepackage{xcolor}

\begin{document}	
\title{A Survey on Universal Design for Fitness Wearable Devices}
\author{Hongjia~Wu, Mengdi~Liu
\thanks{H.Wu (correspondence author) is with SimulaMet and OsloMet, Oslo, Norway. E-mail: \textit{hongjia@simula.no}. M.Liu is with BD Law Firm, Beijing, China. E-mail: \textit{lmd0224@gmail.com}.}
}


\maketitle
\begin{abstract}
Driven by the visions of Internet of Things and 5G communications, recent years have seen a paradigm shift in personal mobile devices, from smartphones towards wearable devices. Wearable devices come in many different forms targeting different application scenarios. Among these, the fitness wearable devices (FWDs) are proven to be one of the forms that intrigue the market and occupy an increasing trend in terms of the market share. Nevertheless, although the fitness wearable devices nowadays are functionally self-contained based on the advanced sensor, computation, and communicative technologies, there is still a large gap to truly satisfy the target customer group, i.e., accessible to and usable by a larger quantity of users. This fuels the research area on applying the universal design principles to fitness wearable devices. In this survey, we first present the background of FWDs and show the acceptance and adaption challenges of the corresponding user groups. We then review the universal design principle and how it and its relative approaches could be used in FWDs. Further, we collect the available FWDs that bear the universal design principles in their development circles. Last, we open up the discussion based on the surveyed literature and provide the insight of potential future work.  
\end{abstract}
\IEEEpeerreviewmaketitle
\section{Introduction}

\IEEEPARstart{W}{earable} devices are defined as worn, user-controllable, and real-time responding devices \cite{mann1997historical}. Specifically, as the natural added values the wearable device can bring upon fitness monitoring and instruction functionalities, the Fitness Wearable Device (FWD) intrigues the commercial market and has gained a new level of popularity. The exploitation of FWDs has been rated as the No.1 fitness trend over the past few years in terms of the market share, as well as one of the segments with the highest growth rate in the horizontal fitness business industries\cite{thompson2018worldwide},\cite{thompson2019worldwide}. FWD can help to track various human daily fitness conditions \cite{krey2020wearable} such as steps, distance, calorie consumption, sleep, and nutrition, based on the multi-modal sensor set such as acceleration sensors, gyroscopes etc. The users can exploit the functionalities of FWDs to their training: expanding services, reaching a clientele far beyond the usual geography, and providing individualized training programs better than ever before \cite{liguori2018fitness}. 

Nevertheless, FWDs do not satisfy the consumers to a degree as they are supposed to be. Studies \cite{ledger2014inside} \cite{tomberg2019designing} have found that many customers abandon their fitness devices very quickly because of various reasons as they are considered not helpful. At the same time, users have kept complaining about particular problems in their wearable devices. What's more, FWDs are still lacking accessibility for people with multiple impairments \cite{malu2016toward}. Therefore, there is still a spate of work required to make FWDs more usable and accessible to more users. 

Meanwhile, Universal Design (UD) attempts to make products, as well as information technology, accessible to and usable by all without regard to gender, ethnicity, health or disability, or other factors that may be pertinent \cite{story1998universal}. By applying universal principles to the design and evaluation process of FWDs, designs can be made accessible to and usable by the broadest range of users with the requirement for the fitness service\cite{preiser2008universal}. 

Over the last few years, a significant volume of scientific articles have been published in the literature on applying UD principles (or similar guidelines) to the design of FWDs. While the existence of a large number of scientific articles, there are few survey papers on this topic. In this paper, we survey the background of FWDs and the challenges encountered when laying in the market. We also give a bird view on the UD and the guidelines on how to apply them in the design and evaluation process of FWDs. Finally, we survey the available FWDs with universally designed features. The paper is organized as follows. 

\begin{itemize}
	\item First, we present the background of FWDs and show the user acceptance and adaption challenges in Section \ref{w}.   
    \item Then, we give a bird view of UD and illustrate why and how UD and its relative approaches could be used in FWDs in Section \ref{UD}.
    \item Next, we depict the available FWDs that introduce the UD principle in their designing process in Section\ref{E}. 
    \item Finally, we open the discussion about the future development trajectory in Section \ref{lesson}. We conclude our work in Section \ref{conclusion}.
\end{itemize}

\section{FWDs and their users study} \label{w}

\subsection{Wearable devices in fitness}

The concept of wearables first appeared in the mid-1990s, when carrying an ‘always-on’ computer combined with a head-mounted display and control interface became a practical possibility. In the “Wearables in 2005” workshop (held in 1996), attendants defined wearable computing as data gathering and disseminating devices that enable the user to operate more efficiently. They also determined that the wearable devices are carried or worn by the user during normal execution of his/her tasks \cite{randell2005wearable}. One of the first advocates of the wearables, Steve Mann, further defined wearable devices as worn, user-controllable, and real-time responding devices \cite{mann1997historical}. 

Today, wearable devices have different types such as smartwatches, wrist bands, smart glasses, etc., and they can sense, collect, upload physiological data in a 24x7 manner \cite{seneviratne2017survey}. Wearable technology such as GPS or inertial measuring unit (composed of accelerometer, gyroscope, and magnetometer) sensors are being implemented to help to monitor and/or stimulate, and/or treat, and/or replace biophysical human function \cite{lymberis2003intelligent},\cite{aroganam2019review}. They are low-cost, independent devices targeting personal use. Because of their small size and high capacity of the processing data, they can be placed almost anywhere on the body. They can be useful to individuals in diverse scenarios such as indoor localization and navigation \cite{yang2015wearables}, \cite{lee2016rgb}, physical and mental health monitoring \cite{vidal2012wearable},\cite{wijsman2011towards}, sport analytics \cite{anzaldo2015wearable}, etc.

Among different applicable scenarios, wearables for fitness have won strong interests from customers. An increasing number of people purchase FWDs. According to researches \cite{thompson2018worldwide}, \cite{thompson2019worldwide}, \cite{niazmand2011quantitative}, the majority of consumer interests in the wearable device lay in the field of fitness and wearable device has been rated as the No.1 fitness trend over the past a few years. FWDs can help to track and monitor daily fitness conditions such as steps, distance, calorie consumption, sleep, and nutrition. In addition, some FWDs also offer position-based tracking using the GPS or motion tracking with acceleration sensors or gyroscopes \cite{krey2020wearable}. They can range from general devices like fitness trackers, body weight and sleep monitors, to fitness performance devices like smart socks to improve running style \cite{hansel2015challenges}. For such added value they can provide for the trainers, \cite{liguori2018fitness} states that the market of FWDs has already maintained its impetus and will continue such a trend, which reaches a clientele far beyond the usual geography but individualized and customized fitness program better than ever before. 

\begin{figure}[t]
	\centering
	{\includegraphics[width=\columnwidth]{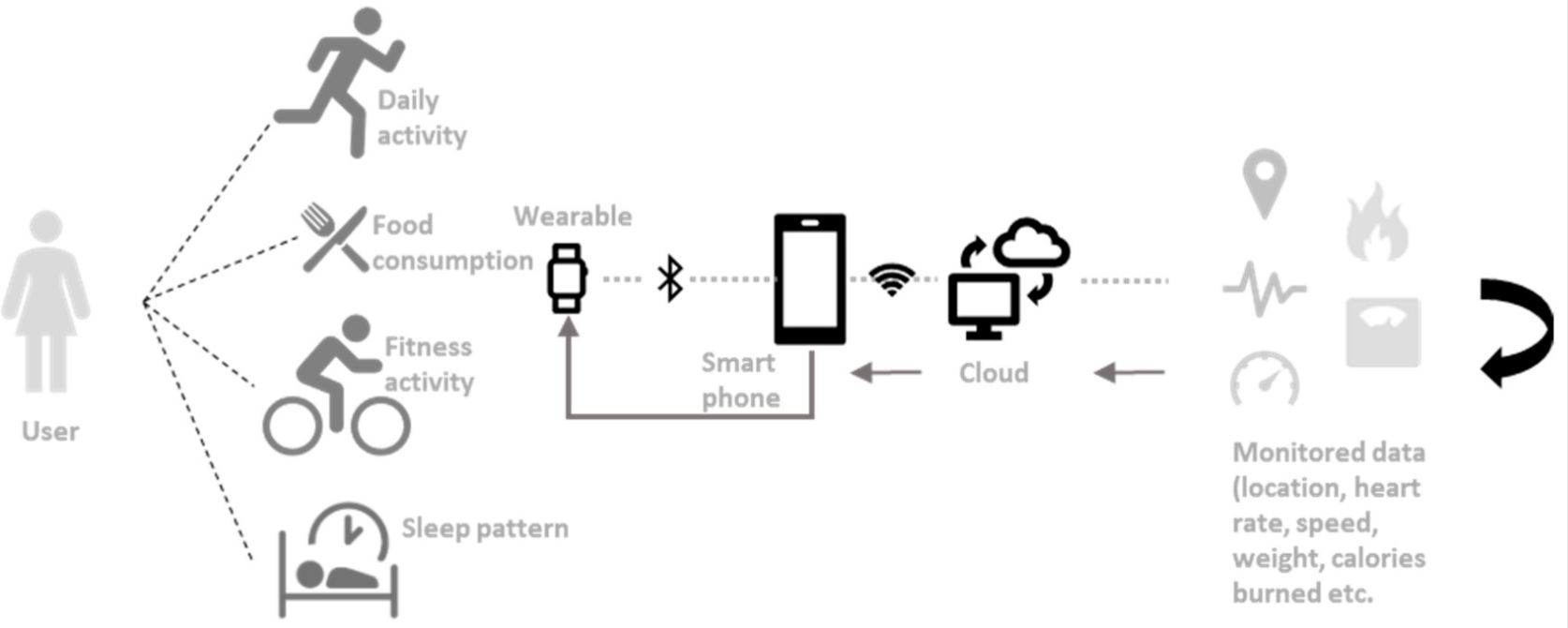}}
	\caption{Block diagram example of a FWD process \cite{aroganam2019review}}
	\label{fig_sensor}
\end{figure}

Figure~\ref{fig_sensor} elaborates on an example of what a current consumer FWD functional process would look like \cite{aroganam2019review}. As shown, it depicts how FWD can be used for lifestyle applications (weight, calories burned, heart rate, speed, etc.). This example explores and exploits the multiple technological factors to give users the intended feedback. The users will have their activity monitored via various sensors passively or actively inputting some of the data by themselves (e.g., food that eaten), which can then be communicated to a smartphone and further to the provider’s cloud service. The data then gets processed, thus they become useful for the user to understand. This is fed back into either the paired smartphone or the wearable itself, depending on the type of display. Further, although it has not occurred yet, it can be foreseen, in the upcoming 5G era, the wearable device will directly communicate with the edge service\cite{mao2017survey} or cloud service, thanks to the mMTC\cite{popovski20185g} slice of the 5G.

\subsection{User study in FWDs}

While the rapid growth of FWDs in the market, they do not satisfy the customers as they are supposed to be. According to \cite{motti2014human} more than half of U.S. consumers who have owned an activity tracker no longer use it. A third of U.S. consumers who have owned one stopped using the device within six months of receiving it. Similarly, studies \cite{ledger2014inside} \cite{tomberg2019designing} have found that many customers abandon their fitness devices because of various reasons, and those devices that are considered not helpful will be abandoned very quickly. In addition, users have kept complaining about particular problems in their wearable devices, including, e.g., poor device autonomy, data validity and robustness, allergy problems \cite{pustiek2015challenges}. In addition, for people with mobility impairments, accessibility challenges are existing in the FWDs. \cite{malu2016toward} found that fitness tracking users have encountered persistent accessibility issues, ranging from the basic form factor of the device to what is tracked. 

Active researches have been conducted to try to find the reasons beneath the abovementioned fact. \cite{wang2015empirical} discovered consumer decision to buy a wearable device relies on the acceptance and adoption of the wearable technology. They found FWD users pay more attention to hedonic motivation, functional congruence, social influence, perceived privacy risk, and perceived vulnerability in their acceptance of wearable technology since they have more perceptions of the enjoyment, comfort, and pricing reasonability of the products. Furthermore, \cite{park2016understanding} showed user adoption of FWDs are depending on usefulness, control, innovativeness, and cost. \cite{kim2015acceptance} examined the user acceptance pattern of FWDs and identified effective quality, relative advantage, mobility, availability, and cost as key psychological determinants of the adoption of wearable devices. Based on it, \cite{lunney2016wearable} claimed FWDs that are perceived as both useful and easy to use are most likely to be adopted by consumers and successful in the quickly growing market. Several studies \cite{lunney2016wearable} \cite{adapa2018factors} \cite{talukder2019acceptance} agreed on the positive effect of social influence on the intention to use FWDs for customers. 

What's more, because of the diversity of lifestyle and requirements of individuals, there is a need for the wearable device be able to be customized easily to fulfill different requirements. \cite{koo2018explorations} compared novice users and experienced users and found different user preferences while using FWDs. Novice users are proven to prefer to track social media posts of others. Experienced users are more reluctant to share their data with others than novice users. \cite{hansel2015challenges} proposed that the elderly may need a different type of support to help trigger health behavior changes, while younger people may consider exercise and diet monitoring function more important while choosing FWDs. FWDs need to support a person’s lifestyle and be able to accessible to and used by each user and meet their requirements. 

In summary, migrating the design of FWDs to be more than functionally self-contained is necessary. By making the design more accessible and usable for more users can be one of the significant factors to improve the users' satisfaction degree. In the following section, we review the UD concepts and its principles and survey the studies regarding how to apply universal design principles to FWDs design to make them more accessible and usable for more users.

\section{Universal Design Theory in FWDs}\label{UD}

In this section, we will introduce the UD, its principles and the design concepts with similar design goals. Then we will review studies on why should designers apply UD to the design of FWDs. Finally, we will survey the academic fundings on UD principles design guidelines on FWDs. 

\subsection{UD and its principles}

Universal Design (UD) attempts to make products, as well as information technology, accessible to and usable by all without regard to gender, ethnicity, health or disability, or other factors that may be pertinent \cite{preiser2008universal}. The origins of UD can go back to the period after World War II when many veterans returned from the war and required rehabilitation in order to continue their normal lives \cite{story1998universal}. Then since 1985, UD has established itself as a potent factor in improving the quality of life for everybody, and on a global basis. UD is now a popular design framework that is used in the different areas of design and development from architecture to product design \cite{tomberg2017universal}. It accommodates the range of variation inherent not just in diverse populations but in varied circumstances, variations in functional characteristics can happen as a result of illness, injury, or aging, or caused by harmful situation and environment \cite{story2001universal}.  

To make UD less complicated in practice, researchers in the Center for UD \cite{story1998universal} assemble a set of principles in 1998, which specify the way designs can be made accessible to the broadest range of users \cite{wentzel2016wearables}. There are seven principles, that could be applied to evaluate the existing designs, guide the design process, and educate designers and consumers about the characteristics of more usable products and environments \cite{connell1997ud}. The seven UD principles with the corresponding descriptions are listed in Table \ref{tab:ud}.

\begin{table}[t]
\caption{Principles for UD and their definitions \cite{connell1997ud}}
\begin{tabular}{|c|l|}
\hline
Principles                          & \multicolumn{1}{c|}{Description}                                                                                                                                                           \\ \hline
Equitable Use                       & \begin{tabular}[c]{@{}l@{}}The design is useful and marketable to peo-\\ ple with diverse abilities\end{tabular}                                                                           \\ \hline
Flexibility in Use                  & \begin{tabular}[c]{@{}l@{}}The design accommodates a wide range of \\ individual preferences and abilities\end{tabular}                                                                    \\ \hline
Simple and Intuitive Use            & \begin{tabular}[c]{@{}l@{}}Use of the design is easy to understand, \\ regardless of the user’s experience, knowle-\\ dge, language skills, or current concentra-\\ tion leve\end{tabular} \\ \hline
Perceptible Information             & \begin{tabular}[c]{@{}l@{}}The design communicates necessary infor-\\ mation effectively to the user, regardless \\ of ambient conditions or the user’s sensory \\ abilities\end{tabular}  \\ \hline
Tolerance for Error                 & \begin{tabular}[c]{@{}l@{}}The design minimizes hazards and the ad-\\ verse consequences of accidental or unin-\\ tended actions\end{tabular}                                              \\ \hline
Low Physical Effort                 & \begin{tabular}[c]{@{}l@{}}The design can be used efficiently and com-\\ fortably and with a minimum of fatigue\end{tabular}                                                               \\ \hline
\begin{tabular}[c]{@{}c@{}}Size and Space for \\ Approach and Use\end{tabular} & \begin{tabular}[c]{@{}l@{}}Appropriate size and space is provided for \\ approach, reach, manipulation, and use re-\\ gardless of user’s body size,posture, or \\ mobility\end{tabular}    \\ \hline
\end{tabular}
\label{tab:ud}
\end{table}

Then in 2007, Erlandson \cite{erlandson2007universal} proposed an extended list of principles with a hierarchical structure, which allows grouping of the principles and establishes relationships between them. All principles were distributed in three main groups: Transcending principles, Process related principles, and Human factors principles (see Figure~\ref{fig_Erlandson}). The UD principles in Erlandson's hierarchical model are shown in Table \ref{Erlandson}.

\begin{figure}[t]
	\centering
	{\includegraphics[width=\columnwidth]{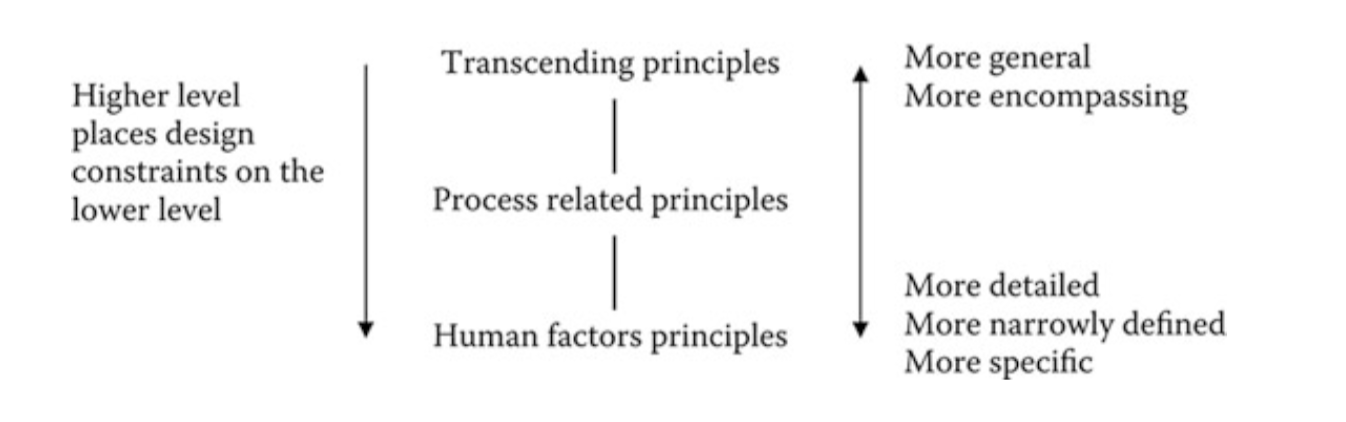}}
	\caption{The hierarchical structure of the UD principles \cite{erlandson2007universal}}
	\label{fig_Erlandson}
\end{figure}

\begin{table}[t]
\caption{UD principles in hierarchical model of Erlandson \cite{erlandson2007universal}}
\centering
\begin{tabular}{|c|c|}
\hline
Levels                                    & Principles               \\ \hline
Transcending principle                    & Equity                   \\ \hline
\multirow{4}{*}{Process principles}       & Flexibility              \\ \cline{2-2} 
                                          & Error management         \\ \cline{2-2} 
                                          & Efficiency               \\ \cline{2-2} 
                                          & Stability/predictability \\ \hline
\multirow{3}{*}{Human factors principles} & Ergonomics               \\ \cline{2-2} 
                                          & Perception               \\ \cline{2-2} 
                                          & Cognition                \\ \hline
\end{tabular}
\label{Erlandson}
\end{table}

Meanwhile, there are several concepts used in different design approaches have similar goals as UD. UD has its roots in the Barrier-free design and accessible design. Accessible design can maximize the number of potential customers by designing products usable by most users without any modification, making products or services adaptable to different users and having standardized interfaces to be compatible with
special products for persons with disabilities \cite{iso2001guidelines}. UD can be used interchangeable with the term design for all \cite{stephanidis2001user}. Design for all \cite{EIDD} aims to enable all people to have equal opportunities to participate in every aspect of society. To achieve this, everything that is designed and made by people to be used by people – must be accessible, convenient for everyone in society to use and responsive to evolving human diversity \cite{balaram2001universal}. Inclusive design bears similarities to UD and design for all, but with the requirement to also include the concept of ‘‘reasonable’’ in the definition \cite{persson2015universal}. It suggests that the inclusion of people with disabilities can be disregarded if considered too difficult to achieve \cite{tbsi2005design}. Human-centered design approach is taken, to ensure the design guidelines match both designer and developer needs, as well as VIP needs and wishes in wearable technology solutions \cite{wentzel2016wearables}. Based on it, researches \cite{newell2000user} have suggested to adapt User-Sensitive Inclusive Design (USID) as an extension of User-Centred Design (UCD) because the wide variety of functionality and characteristics of user groups is very hard to be represented by a small sample. 

\subsection{What will FWDs benefit from UD}

Many pieces of research have been conducted to explore why should FWDs be designed universally. While developing the UD principles, \cite{story1998universal} claimes UD principles could be applied to evaluate existing designs and guide the design process. Similarly, \cite{connell1997ud} proposes designers who use UD principles can create solutions that are usable by as many people as possible instead of trying to create special solutions for a specific disability. \cite{balaram2001universal} further proposes that applying UD principles, there will be four major benefits: Educating for the future; Positive thinking by user groups; Increasing the usability range; Bridging the gap between people. 

The extended list Erlandson \cite{erlandson2007universal} proposes simplified the practical implementation of UD principles. It is important because it explicitly adds the transcending and human-factor-related consideration to the design process. Even no product can truly fulfill the needs of all users, if all users’ needs are taken into consideration in the initial design process, the result is a product that can be used by the broadest spectrum of users. 

\cite{motti2014human} first points out human factors need to be addressed from the early design stage of wearable applications. By excluding the users’ perspective during the design phase, the devices’ acceptance is likely to be compromised, invasive or cumbersome and especially if recordings must be continuously made in their natural environments. But there are several trade-offs in the FWD. \cite{angelini2013designing} proposes that by focusing on the feasibility of an individual approach, often usability and wearability are neglected; when excluding the users’ perspective during the design phase, the devices’ acceptance is likely to be compromised, especially when the resulting device is bulky. 

Studies \cite{ferraro2011designing} \cite{frances2018role} conclude the creation of the FWDs requires specific concepts, electronics, and software that consider the diversity of potential users and their environments. Thus, successful wearable usability is no longer about providing technical success, but rather about creating an optimal user experience.

\cite{tomberg2019designing} agrees it is not sufficient to just apply the usability rules while designing wearable devices, because the nature of user interfaces for FWDs is versatile. Instead, the human factor design principles should be applied for use in the design and evaluation of wearable devices. With increase accessibility, wearable devices open up information support and assistance to persons with a disability and can stimulate empowerment and participation.

Retail marketing of inclusive designs has the potential to significantly impact the availability and affordability of wearables for people with disabilities. \cite{aroganam2019review} reviews the FWDs used in consumer sports applications and found having a user-centered design process helps build trust amongst sports wearable tech consumers. The do-it-yourself and maker movements are also impacting the way devices are designed and obtained as well as empowering end-users and members of the broader community \cite{awori2017maker}. In ergonomic and anthropometric terms, positioning of the wearables is crucial to success. Flexible sensors can be practical if the design of the fitness wearable is ergonomic enough for the user to easily adapt to the change \cite{aroganam2019review}.

\subsection{How to universally design FWDs}

We survey the academic findings that have outlined the principles and guidelines on how to universally design FWDs, and that have focused on important factors that should be considered for designing FWDs more universally accepted in this subsection. 

\cite{suhonen2012haptically} raises insight about people’s preferences of FWDs related to the usage styles and use purposes of the devices. Then \cite{motti2014human} outlines 20 human-centered principles to be considered during the design of FWDs. The 20 principles identified with the systematic literature review include Aesthetics, Affordance, Comfort, Contextual-awareness, Customization, Ease of Use, Ergonomy, Fashion, Intuitiveness, Obtrusiveness, Overload, Privacy, Reliability, Resistance, Responsiveness, Satisfaction, Simplicity, Subtlety, User-friendliness, and Wearability. They also explained how each principle can be incorporated during the design phase of the wearable device creation process and prove that user-centered approaches are useful in this process, alternating iterations, and evaluations, such as focus groups, interviews, and surveys.

\cite{tomberg2015applying} examines UD principles to FWDs. For the Equitable Use principle, it should be tested when the design process starts. It can help to understand if the device idea contradicts to equality and may provide prompts for use of the same FWDs by diverse groups of people. Flexibility in Use is the process-related principle, which should be applied to the design of wearable devices to provide choice in methods of use, facilitate the user’s accuracy and precision, and provide adaptability to the user’s pace. For Tolerance for Error, if the fitness wearable is part of the Synced Lifestyle, then the solution should allow for synchronization errors and corrections. A simple and intuitive use principle is another process-related principle. Any functions that can be enabled, should be disabled easily if the users wish to shut it down. FWDs are required to work seamlessly with the wearer, which can only be accomplished if the wearable is easy to learn and simple to use. The aim of Perceptible Information is to communicate necessary information effectively to the user, regardless of ambient conditions or the user’s sensory abilities. Low Physical Effort should be taken while designing FWDs to be worn for extended periods of time, they need to have a low amount of physical effort involved in their use. Size and Space for Approach and Use promotes an appropriate size and space, which should be provided for approach, reach, manipulation, and use regardless of user’s body size, posture, or mobility.

The extended lists \cite{erlandson2007universal} proposed, have been applied to the design of FWDs. While Erlandson put equitability as a prominent principle that integrates other UD principles, when applying it to the design of FWDs, \cite{tomberg2019designing} proposes to adapt equity, on the base of the top place in Erlandson’s hierarchy, as a preliminary step to support the positioning of a project’s compliance to the UD principles. It further integrated all three layers of the Erlandson’s model in an iterative cycle and showed the layers can proceed during evaluation (see Figure~\ref{fig_Erlandson}). After starting with transcending principle and using it with a “compass,” one goes into the process-related principles where top-level principles are restricting her. Next, moving to human factors principles and being again restricted by top-level principles, she goes back to the transcending principle with new knowledge and starts a new iteration. This way equity places higher demands on designers studying actual users and working with them as an integral part of the whole design process. 

\begin{figure}[t]
	\centering
	{\includegraphics[width=.7\columnwidth]{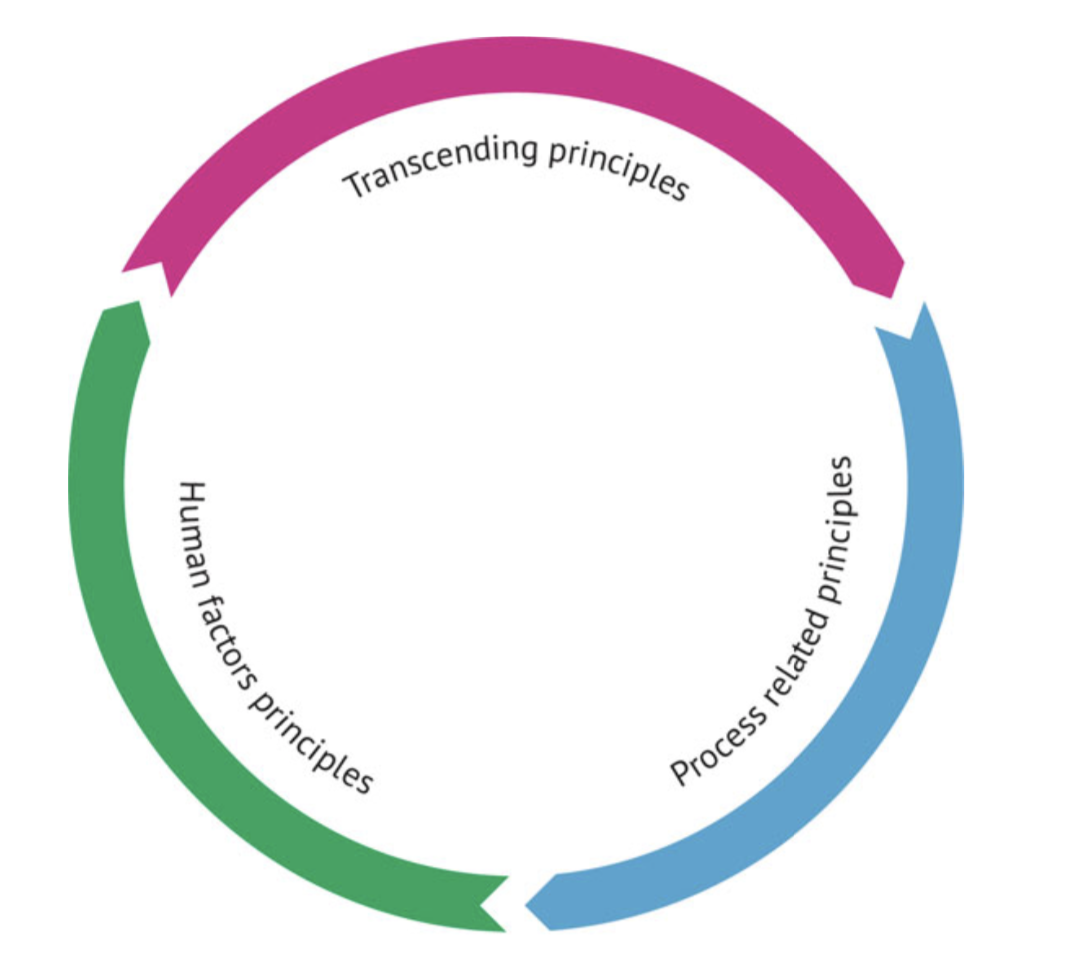}}
	\caption{Iterative workflow adapted from Erlandson model \cite{tomberg2019designing}}
	\label{fig_Erlandson}
\end{figure}

The process-related principles are constrained by the transcending principles and at the same time provide the constraints to the human factors principles, those principles (listed in Table \ref{Erlandson}) are associated with the usability of a product. Usability heuristics \cite{nielsen199510}, which cover all UD process-related principles, are followed by designers. Human factors principles are followed bu using Inclusive Design Toolkit (IDT) to assess human factors. It can place demands on the users’ capabilities. Because human factors defined in IDT are well aligned with Erlandson’s human factors principles, using the IDT scale is especially useful when designers develop a set of personas, based on the actual encounters with target user groups with the different abilities.

\cite{wentzel2016wearables} describes guidelines to develop accessible FWDs. Use multimodal presentation of information for users with different preferences and abilities; Use multimodal interaction to allow users to interact with a system following their individual preferences and suited to their personal needs. A third principle is then added to address human feedback: The FWDs should provide relevant feedback on the user behavior and the system actions, which can consist of positive confirmation and reinforcement of actions, or notification and instructions on unexpected or incorrect behavior or actions. To accommodate the changing preferences of users in various settings, the system settings should be adaptive and/or adaptable. Adaptation of preferred settings (e.g., for input/output modalities, feedback intensity) should be contextual; based on localization, task, and/or user preferences. It is advised that the FWDs should be self-learning to enable optimal automated adaptive settings.

\begin{figure}[t]
	\centering
	{\includegraphics[width=.95\columnwidth]{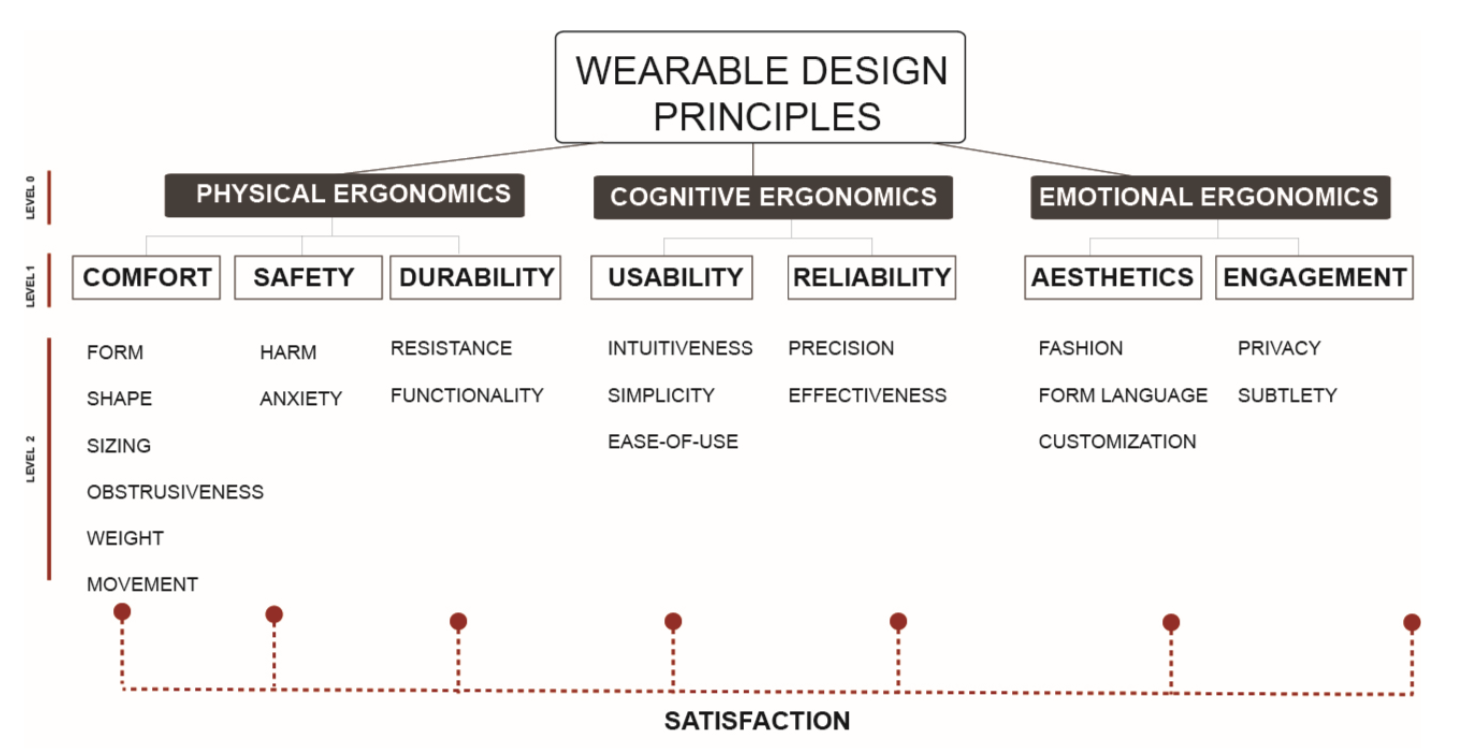}}
	\caption{Wearable design requirements map \cite{frances2018role}}
	\label{fig_map}
\end{figure}

Based on the quantitative and qualitative analysis of the literature reviews, \cite{frances2018role} proposed a multi-level framework that organizes and relates design criteria shown in Figure~\ref{fig_map}. Level 0 of this framework includes the three ergonomic domains (physical, cognitive, emotional). In Level 1, the requirements have a more general character as Comfort, Safety, Durability, Usability, Reliability, Aesthetics and Engagement. Finally, in Level 2 the requirements are divided into more detailed requirements, which can be measured with different methods, whether quantitative or qualitative. Moreover, in this framework, some requirements from the studies that have different names but similar meanings are unified; for example,  “ease of use” and “intuitiveness”. Following this framework, designers will aim to be a first step in the building of a solid framework for the user-centered design of wearable devices, which will be further complemented and discussed with other researcher’s point of view and experience.

\cite{bernal2017safety++} focused on a typical User-Centered Design (UCD) Approach, composed of four phases: discover, define, develop, and deliver, to study how the UCD approach applied to the design of new FWD for safety in the energy industry. The first step of the project (discover) consisted in performing research and analysis on the user and the context, In the second step of the process (define), researchers identified four main problems we intended to address by designing a new IoT system for the energy industry: Exposure, Man down, Fall from height and Load lifting. Once the problems to address were clarified, the research team starts the development phase, consisting of designing new solutions aiming to solve the mentioned safety issues. 

\cite{harte2017human} proposes a methodology based on human factors to design wearable devices. They used the three-phase approach to develop a more friendly wearable design, Setting a base; Design requirements classification; Requirements analysis which has two steps: a. Quantitative analysis: different parameters were defined as quantifying parameters. b. Qualitative analysis: To establish connections between different requirements, some user-centered design tools were used.

\begin{figure}[t]
	\centering
	{\includegraphics[width=.85\columnwidth]{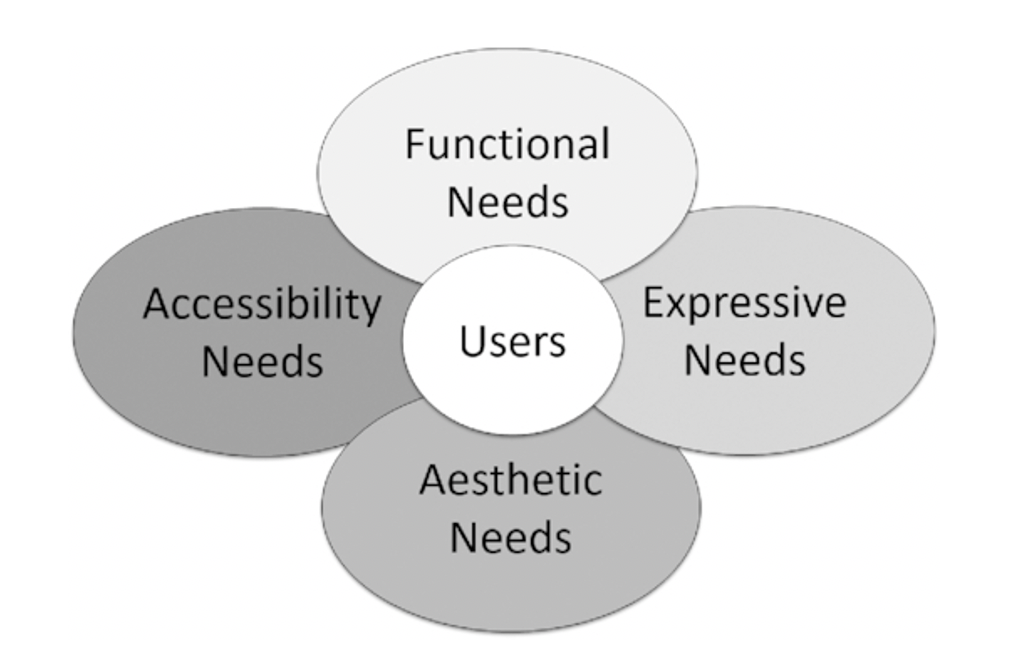}}
	\caption{The FEA2 model  \cite{hall2018design}}
	\label{fig_ffa}
	\vspace{-5mm}
\end{figure}

The FEA2 (Function, Expressiveness, Aesthetics, Accessibility) model is recently developed in Dr. Lobo \cite{hall2018design}, to guide the design of FWDs for disabled people. The model is an extension of the existing FEA apparel design user needs model integrated with the design approach of engineering and the philosophy of patient-centered health care. The FEA2 model incorporates user needs related to function, expressiveness, aesthetics, and accessibility (see Figure~\ref{fig_ffa}). The FEA2 multidisciplinary framework presents a model for developing innovative and impactful rehabilitation devices for fitness in the real world.

For designing FWDs for the special group of people, \cite{khakurel2018crafting} presents a series of guidelines for integrating usability into the design of FWDs targeted to older adults. For enhancing Usability for Hardware, Consider Configure-to-Order (CTO) Products (with a wide device and screen size measurement and shape variations); Consider Maximum Magnitude of Effect for Minimal Means (For example, removing unwanted hardware); Consider Improving Sensor Precision and Consider Culture While Designing the Devices are key elements to improve the hardware usability for elderly people. Consider Alternative UIs, support User personalization, support users' requirements. 

\cite{lowens2015design} proposes design Recommendations to improve user interaction with wrist-worn devices for fitness and health. These recommendations include: Device dimensions, being compact, light, and comfortable, minimalist and unobtrusive; Durability of the display, needs to be resistant to be worn daily; Context-awareness. The device, more efficient algorithms to properly detect the movements and react accordingly based on user context; Accuracy of data collection; Definition of interfaces and interaction. Improving user interface and employing a more efficient display. Availability of visualization tools. Offer solutions to analyze user performance and activity level while offering tools to interpret data; Ease of use. The gestures should be intuitive, i.e. ease to learn and to remember but still detected accurately.

\cite{ferraro2011designing} proposes a user-needs driven design methodology is proposed that promotes collaborative design with users. It addresses a breadth of technical, functional, physiological, social, cultural and aesthetic considerations that impinge on the design of clothing with embedded technologies, that is intended to be attractive, comfortable and fit for purpose for the identified customer. To aid decision-making, the design process requires an overview of the profile of the target customer in terms of gender, age group, and an indication of the proposed category of smart textile products to be developed.  Concerns social and cultural issues, historic context and tradition, corporate and work culture, participation patterns and levels, status, demographics, and the general health and fitness of the wearer will impinge on the design of smart clothes and wearable technology. An investigation of the lifestyle demands of the wearer, in terms of behavior, environment and peer group pressure, is needed to provide an awareness of both clothing requirements and the application of emerging wearable technologies that have appropriate functionality and true usability for the identified user.

\section{Case study on FWDs} \label{E}

We have surveyed the theoretical studies of UD guidelines and principles in the field of FWDs. In this section, we look at the available FWDs that have introduced UD and its similar concepts in two subsections, a group of studies towards the increase of usablility for more users and a group of studies towards that the increase of accessibility for more users.   

\subsection{More usable FWDs for more users}

Considering UD in fitness wearable design, FWDs can enable better user experience with user feedback. For example, they can provide real-time and post-workout feedback, with no user-specific training and no intervention during a workout \cite{morris2014recofit}. They can provide feedback about amounts of activity performance that can be used to motivate users to improve their endurance and physiological and mental health \cite{cadmus2015randomized}. They can also give continuous haptic and visual feedback to the user for enhanced user experience without any prior training \cite{pruthi2015maxxyt}.

MOPET \cite{buttussi2008mopet} as a context-aware and user-adaptive wearable system supervises a physical fitness activity based on alternating jogging and fitness exercises in outdoor environments. By exploiting real-time data coming from sensors, knowledge elicited from a sports physiologist and a professional trainer, and a user model that is built and periodically updated through a guided autotest, MOPET can provide motivation as well as safety and health advice, adapted to the user and the context. To better interact with the user, MOPET also displays a 3D embodied agent that speaks, suggests stretching or strengthening exercises according to the user’s current condition, and demonstrates how to correctly perform exercises with interactive 3D animations. 

Battery-less technology can be developed for lightweight and small-sized (as batteries tend to be heavy) purpose, and for a more sustainable solution (better energy conservation). Using kinetic energy (body movement) to charge the wearable can help to design light and unobtrusive devices.  \cite{wu2015cellular} applies cellular PP piezoelectrics for harvesting human kinematic energy and detecting human physiological signals. \cite{magno2016kinetic} evaluates and integrates a highly-efficient kinetic harvester circuit to power autonomous wearable devices, exploiting the energy gathered from human motion. With the advancement of wireless charging, this can also be used for design considerations where the components can be embedded into equipment for easy maintenance \cite{aroganam2019review}.

Material advancements, conductive threads, and smart material being embedded into clothing can make the FWDs more comfortable and common in the commercial market. Stretch sensors made from conductive threads using a cover stitch \cite{gioberto2012theory}, or using a zig-zag stitch \cite{greenspan2018development}, are being developed and tested to measure movement. These innovative FWDs are designed more comfortable and appealing to users, as user-friendly alternatives to hard, bulky sensors. 

\begin{figure}[t]
	\centering
	{\includegraphics[width=.75\columnwidth]{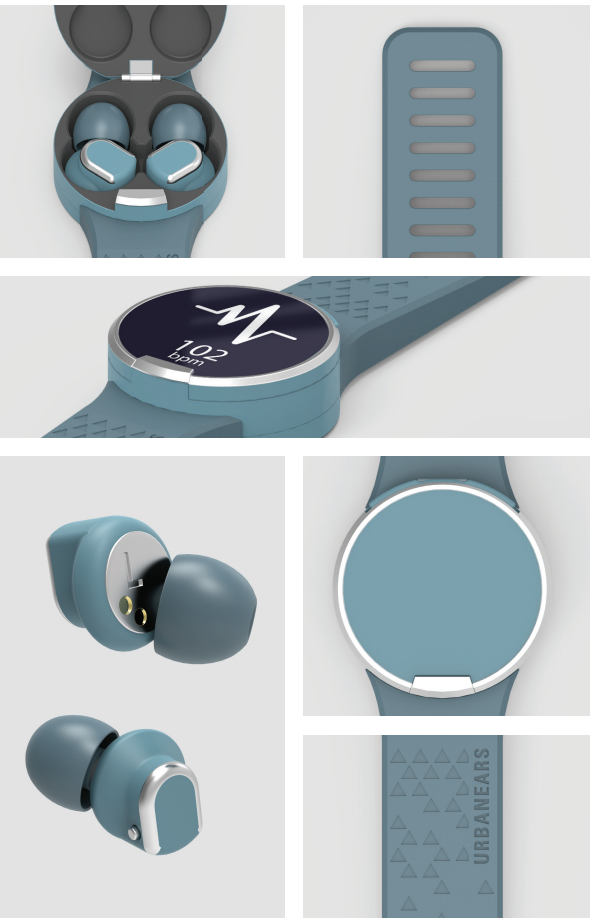}}
	\caption{Urbanears Ugglan \cite{ahmed2018urbanears}}
	\label{fig_kth}
\end{figure}

Having a user-centered design process can win customers' trust. Zepp Play \cite{zepp} uses different combinations of sensors, the same sensor programmed to produce data for different attributes, for different applications as football and golf. Urbanears Ugglan (see Figure~\ref{fig_kth}), \cite{ahmed2018urbanears} adopts a user-centered design approach, to provide a solution to take control of human physical activities. It consists of two parts: true wireless earphones and a wearable wrist accessory that stores and charges the earphones on the go, while also functioning as a fitness tracker. Through user studies, they understood issues users have with their headphones and important requirements from targeted users. Urbanears Ugglancan helps active users who value independence, simplicity, and the ease of use smartly take control of their activities. It has a large range of selection for users to enable or disable the functions easily, which can be seen as a good example of user-centered design.

\cite{zhou2019applying} follows a user-centered approach to create a new mobile app (PittPHR) for fitness trackers by a questionnaire study for needs assessment, there were 609 participants in total. Based on the needs assessment findings, PittPHR is created with different modules as health records, trackers and resources. This app allows users to customize the trackers according to their needs. The usability study participants (17 people) expressed high satisfaction with the product. 

A fitness wearable vast safety++ \cite{bernal2017safety++} is designed and tested with a UCD approach. They focused on the actual user experience in real environments to develop FWDs able to respond to real needs in the field of safety. Their approach focuses not only on the functions and performance of the system but also on transforming connectivity into meaningful communication of risks and incorrect behavior to teammates and supervisors. Their approach minimizes the problems of overconfidence, distraction, ‘macho’ culture and lack of awareness, which represent some of the major causes of incidents in their training.

\subsection{More accessible FWDs for more users}

As of yet, FWDs are still lacked accessibility for disabled people. Among all commercial FWDs, The Apple Watch \cite{apple} is the only mainstream step-counting product that has a mode to measure activity while rolling, which can support step measuring for the users who have trouble in normal walking. But there are still some innovative FWDs that are designed with the consideration of high accessibility in the lab phase.

\begin{figure}[t]
	\centering
	{\includegraphics[width=.85\columnwidth]{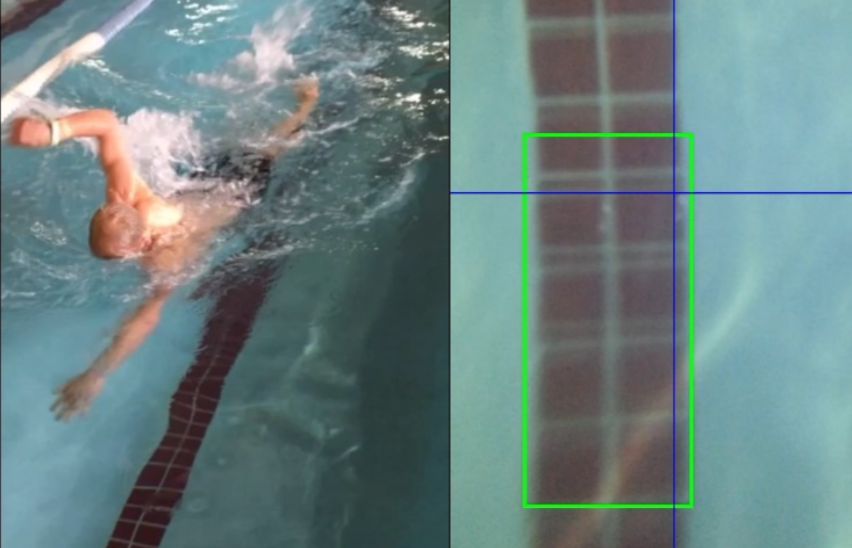}}
	\caption{Goby \cite{muehlbradt2017goby}}
	\label{fig_goby}
\end{figure}

Goby, \cite{muehlbradt2017goby} a swimming aid as shown in Figure \ref{fig_goby}, can provide audio feedback for blind and visually impaired athletes addressed the accessibility of swimming for blind and visually impaired people. Goby’s activity tracker could be worn on the thigh and uses a downward-facing camera to track the swimmer’s position in the pool. Goby detects when the user is swimming outside the lane or approaching a wall and warns the user via an audio notification. 

Playskin Lift \cite{lobo2016playskin} promises assistive and rehabilitative devices for a variety of populations, which are developed and tested of an exoskeletal garment to assist upper extremity mobility and function. Their results suggest that by considering the broad needs of users, including cost, accessibility, comfort, aesthetics, and function, inexpensive wearable devices that families and clinicians can potentially fabricate in their communities to improve function, participation, exploration, and learning for children with disabilities can be designed. 

\begin{figure}[h]
	\centering
	{\includegraphics[width=.8\columnwidth]{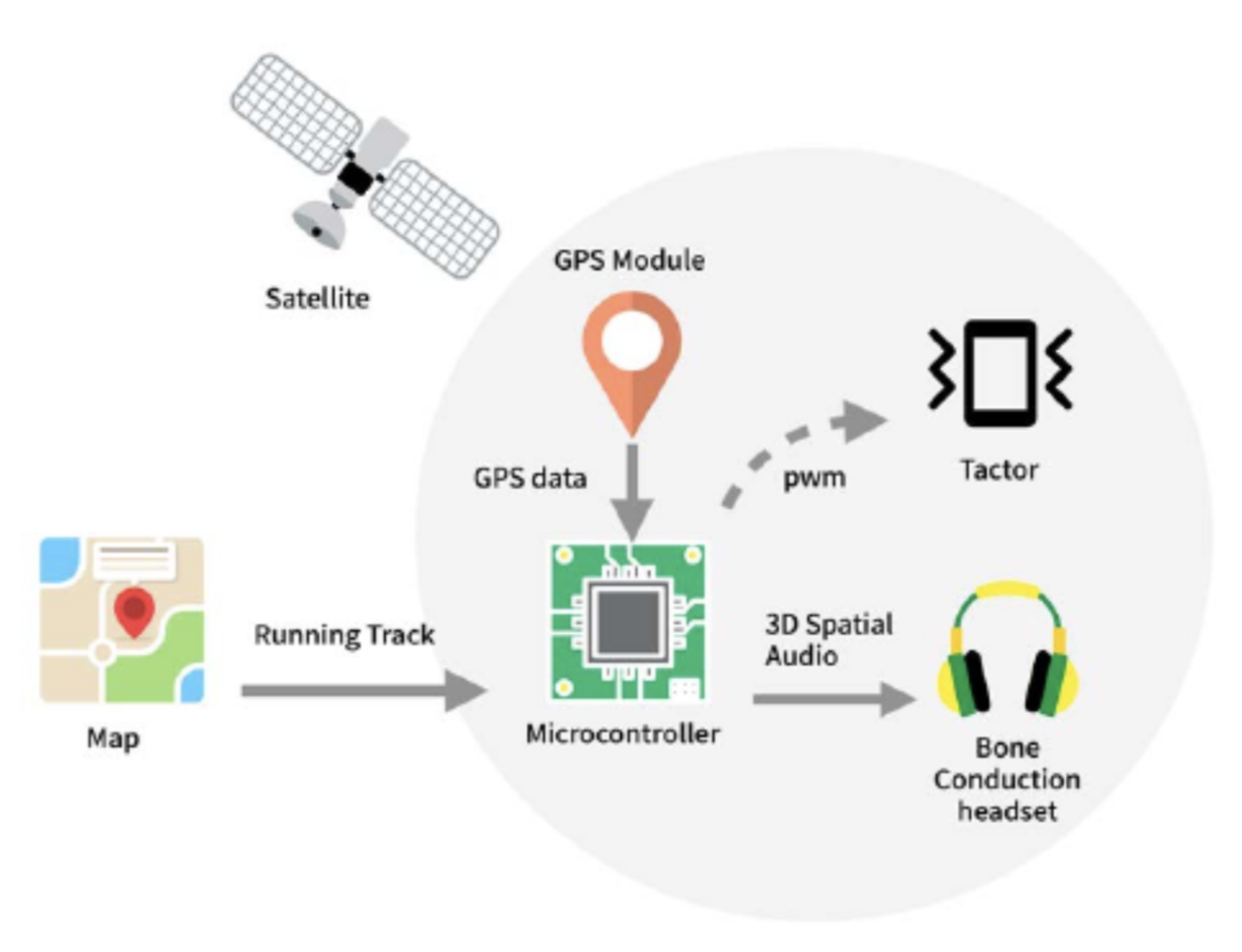}}
	\caption{Marathon Navigation System \cite{zhu2019running}}
	\label{fig_run}
\end{figure}

\cite{zhu2019running} creates a running guide, a pair of wearable glasses based on the insights from user research, enabling visually impaired runners to navigate the running route and finish the marathon more independently and confidently, as shown in Figure \ref{fig_run}.

\cite{kelly2019advancement} is a wearable device concept, that utilizes wearable technology to assist in the development and maintenance of orthopedic health, while avoiding injuries during training. In situations where individuals may be unable to feel or touch while being unable to speak, the energy wristband (NRG) will be programmed with an artificial intelligence-based algorithm that counts eye blinks, as an extension of the personal assistant. With this, individuals with amputated limbs where they may not have a wrist on one hand may still have accessibility using the NRG’s blink algorithm as commands. 

\begin{figure}[t]
	\centering
	{\includegraphics[width=.8\columnwidth]{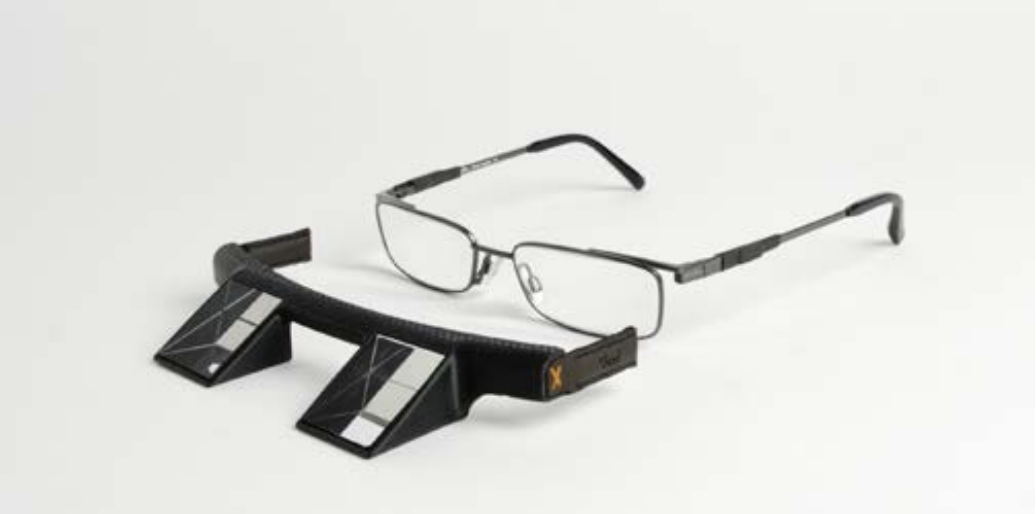}}
	\caption{The reversed prismglasses design \cite{de2012expect}}
	\label{fig_glass}
\end{figure}

\cite{de2012expect} proposes a reversed prism glasses design for ankylosing spondylitis people who cannot lift his head entirely upwards, based on the design process of distinction creation and distinction destruction occupational therapist, professional nondesigners, caregivers and disabled people. Through the design activities within community-based rehabilitation contexts, the reversed prism glasses design in Figure~\ref{fig_glass} can help users ease a lot of friction with some fitness activities. It is compatible with users‘ large collection of eyeglasses, has hands-free aspects, which enable increased usability and user experience.

\begin{figure}[t]
	\centering
	{\includegraphics[width=.65\columnwidth]{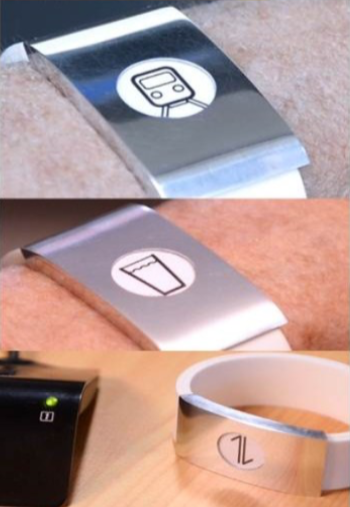}}
	\caption{Functions of the smart bracelet \cite{angelini2013designing}}
	\label{fig_old}
\vspace{-5mm}
\end{figure}

\cite{trujillo2017development} has used wearable sensors to capture the variability and repertoire of limb movements that infants produce across full days. \cite{angelini2013designing} presents the design process of a smart bracelet that aims at enhancing the fitness life of elderly people. The bracelet, in Figure~\ref{fig_old} acts as a personal assistant during the user’s everyday life, monitoring the health status and alerting him or her about abnormal conditions, reminding medications and facilitating the everyday life in many outdoor and indoor activities. Following a user-centered approach, the design process involved several iterations and evaluation of prototypes with older adults. The preliminary interviews will be used to refine the product design to create a new mockup that will be used to conduct a usability evaluation. 

\section{Discussion} \label{lesson}

Based on our survey work on UD for FWDs, we perform the discussion from the following aspects. 

The cutting edge technologies such as material advancements, battery-less technologies, and fast data-processing approaches can enable FWDs towards higher usability. According to our review, the size and weight of the FWDs are among the first considerations for customers on whether to choose. They are normally correlated with the material advancements. Long battery life and fast response of the device are also the key characteristics that customers are concerned with. Therefore, product designers should focus on state-of-the-art technologies that can be applied to the design process to make the FWDs with higher usability for more users. However, those features are not the characteristics that the product designers can easily improve. Most likely, the development parties of those features are domain experts who can help to enable smaller PCB and battery, such as electronic and material scientists. The limitations of technology can be a big obstacle to prevent designing FWDs universally. Therefore, future research can work on technology improvement via tight collaboration to make the FWDs more usable and acceptable for more people. 

While several studies have proposed more accessible FWDs, the accessibility of the FWDs is still a challenging issue. Most devices are composed of neither features that are equipped with strong accessibility characteristics for users nor features that would consider health complications and disability \cite{ahmed2018urbanears}. For instance, as we stated before, current consumer wearable devices that are available cannot meet the needs of the diverse population of wheelchair athletes. More inclusive designs are needed to deliver physical activity information to a more diverse audience of users \cite{carrington2015but}. We also cannot find much work on the FWDs in adaptive sports or for individual adaptive sports athletes. Hence, further work should be done to understand impaired users' requirements which can be a significant step toward a more inclusive fitness community.

The number of researches on applying user-centered design methods for FWDs into practice is limited. When we surveyed, we found most of the researches focus on the guidelines for designing and evaluating FWDs, but few focus on the system studies and experimental evaluations. While the argumentations over general guidelines are helpful, applying those guidelines into practical studies may be more convincing. Through the user case studies, how UD principles can benefit the FWDs design can be showed in more detail. Further, this can form a positive and iterative development loop between the theory and the system design. Nevertheless, we have to admit that organizing such a study usually requires richer resources, including extra funding on hardware in loop design, extra human resources on physical system development, etc. 

Currently, designers know their users through customer studies. This step fundamentally defines the overall development trajectory. Thus, designers should dig deeper to understand the meaningful information based on effective data analytics tool with high efficiency.  Besides depth, the coverage range is also important, as it is directly related to the basic definition of UD. However, to obtain a larger coverage range is equally difficult, considering the availability and diversity of the user data. The crowdsourcing approach seems to be a potential approach to explicitly increase the coverage range and implicitly aids the designers to enhance the analysis depth. Fortunately, the feasibility of this approach also increases a lot with the upcoming internet of things era and the well-recognized data-driven trend in society.

Several topics are not covered in this survey. First, UD for FWDs on the software engineering aspect is not reviewed in this paper. Second, data security and privacy could be a rising issue. The proper use of data is an ethical concern, which can also be considered as a limitation towards FWDs with a higher quantity of users.  Last, another issue faced in advanced smart wearables is the safety of the products. Since the popularity growth in the use of FWDs, safety issues such as sports injuries under improper instructions from  FWDs may happen. 

\vspace{-5mm}
 
\section{Conclusion}
\label{conclusion}
This article has presented a survey of how the UD design principles can be used to aid to increase the satisfaction degree of users, specifically via improving the accessibility and usability for more users. The solutions have been categorized into two types, theoretical proposals, and system work. Further, we discuss the existing solutions and open up the research issues for the future.

\bibliographystyle{IEEEtran}
\bibliography{bibliography_multipath}

\end{document}